UDK O35, R11.

*SZABOLCS NAGY*

# DIGITAL ECONOMY AND SOCIETY – A CROSS COUNTRY COMPARISON OF HUNGARY AND UKRAINE

У даній роботі я вперше проаналізував стан цифрової економіки та суспільства в Угорщині, потім порівняв його з Україною та зробив висновки щодо майбутніх тенденцій розвитку. Використовуючи вторинні дані, надані Європейською комісією, я досліджував п'ять компонентів Індексу цифрової економіки та суспільства Угорщини. Я провів аналіз крос-країни, щоб з'ясувати суттєві відмінності між Україною та Угорщиною щодо доступу до Інтернету та використання пристроїв, включаючи смартфони, комп'ютери та планшети. Виходячи з моїх висновків, я зробив висновок, що Угорщина більш розвинена з точки зору значущих параметрів цифрової економіки та суспільства, ніж Україна, але навіть Угорщина є новою цифровою нацією. Враховуючи високі темпи зростання інтернет-прояву планшета та смартфонів в обох країнах, я очікую більш швидкого прогресу у розвитку цифрової економіки та суспільства в Угорщині та Україні.

**Ключові слова:** цифрова економіка, цифрове суспільство, DESI, Інтернет, використання пристроїв, Угорщина, Україна, порівняння між країнами

В этой статье я сначала проанализировал состояние цифровой экономики и общества в Венгрии, затем сравнил ее с Украиной и сделал выводы относительно будущих тенденций развития. Используя вторичные данные, представленные Европейской комиссией, я исследовал пять компонентов индекса цифровой экономики и общества Венгрии. Я провел кросс-анализ, чтобы выяснить существенные различия между Украиной и Венгрией в плане доступа к Интернету и использованию устройств, включая смартфоны, компьютеры и планшеты. Основываясь на моих выводах, я пришел к выводу, что Венгрия более развита с точки зрения значительных параметров цифровой экономики и общества, чем Украина, но даже Венгрия является новой цифровой нацией. Учитывая высокие темпы роста проникновения Интернета, планшета и смартфонов в обеих странах, я ожидаю более быстрый прогресс в развитии цифровой экономики и общества в Венгрии и Украине.

**Ключевые слова:** цифровая экономика, цифровое общество, DESI, интернет, использование устройств, Венгрия, Украина, сравнение по странам

We live in the Digital Age in which both economy and society have been transforming significantly. The Internet and the connected digital devices are inseparable parts of our daily life and the engine of the economic growth. In this paper, first I analyzed the status of digital economy and society in Hungary, then compared it with Ukraine and made conclusions regarding the future development tendencies. Using secondary data provided by the European Commission I investigated the five components of the Digital Economy and Society Index of Hungary. I performed cross country analysis to find out the significant differences between Ukraine and Hungary in terms of access to the Internet and device use including smartphones, computers and tablets. Based on my findings, I concluded that Hungary is more developed in terms of the significant parameters of the digital economy and society than Ukraine, but even Hungary is an emerging digital nation. Considering the high growth rate of Internet, tablet and smartphone penetration in both countries, I expect faster progress in the development of the digital economy and society in Hungary and Ukraine.

## 1. INTRODUCTION

With an increasing number of people doing a lot of things online and live their life in the virtual world, it is sure that now we live in the Digital Age in which Internet plays a vital role. Growth of Information and Communications Technology (ICT) use has increased significantly over the past three decades (Chavanne, Schinella, Marquet, Frangi & Le Masson, 2015).

The digital society and economy, which is still developing, are totally different from the traditional models. The sharing economy phenomenon is not a temporary trend anymore and has the potential to drastically change competition across the globe (Parente, Geleilate & Rong, 2017).

Internet use also has a favorable effect on both financial development and trade openness. Although increased ICT use might cause higher energy use, Salahuddin, Alam & Ozturk (2016) found that OECD countries can boost their Internet usage without being significantly concerned about its environmental consequences.

Tapscott & Williams (2006) identified four basic principles of the Internet-centric economy as follows: openness, peering, sharing, and acting globally. Internet has a serious impact even on our well-being in four areas. It changes time use patterns, creates new activities, facilitates access to information, and acts as powerful communication tool (Castellacci & Tveito, 2017).

High penetration of Internet, and high rate of digital device usage are the prerequisites of the digital economy and society. Hungary is still an emerging country in terms of digital development.

The main objectives of this paper are

1) to analyze the status of digital economy and society in Hungary,

2) to compare it with Ukraine and

3) to make conclusions regarding the future development tendencies.

## 2. METHODOLOGY

In order to analyze the status of digital economy and society in Hungary, I used secondary data provided by the European Comission (EC, 2017) and the Consumer Barometer with Google (CB, 2017a). First, I have investigated the Digital Economy and Society Index (DESI), which is an online tool to measure the progress of EU Member States towards a digital economy and society. It is a composite index measuring progress in digital through five components. *Connectivity* component includes fixed broadband, mobile broadband, broadband speed and prices. *Human Capital* component measures basic skills and Internet use, advanced skills and development. *Use of Internet* is made up of citizens' use of content as well as communication and online transactions. The *Integration of Digital Technology* is a component in which business digitisation and eCommerce are measured. *Digital Public Services* contains eGovernment (EC, 2017).

The Consumer Barometer is a very powerful tool to help you understand how people use the Internet across the world. "Data in the Consumer Barometer is pulled from two sources - the core Consumer Barometer questionnaire, which is focused on the adult online population, and Connected Consumer Study, which seeks to enumerate the total adult population and is used to weight the Consumer Barometer results" – says the Consumer Barometer about the methodology (CB, 2017a).

### 3. ANALYSIS OF DIGITAL ECONOMY AND SOCIETY

DESI scores range from 0 to 1, the higher the score the better the country performance is. Table 1 shows that the Digital Economy and Society Index in Hungary in 2017 scored 0.46 out of 1.00, which is a slight improvement from 0.43 points in 2016.

**Table 1 DESI scores Source: EC (2017)**

|  | Hungary rank | Hungary score | Cluster score | EU score |
|---|---|---|---|---|
| DESI 2017 | 21 | 0.46 | 0.41 | 0.52 |
| DESI 2016[1] | 20 | 0.43 | 0.38 | 0.49 |

Hungary ranks only 21st after Slovakia and before Cyprus in DESI 2017 out of the 28 EU Member States, sliding back one position in the ranking compared to the previous year. (Figure 1). Therefore, Hungary belongs to the cluster of low performing countries. The best performing country in this ranking is Denmark (DK), followed by Finland (FI), Sweden (SE) and the Netherlands (NL). At the other extreme, the worst performing countries are Romania (RO), Bulgaria (BG) and Greece (EL)

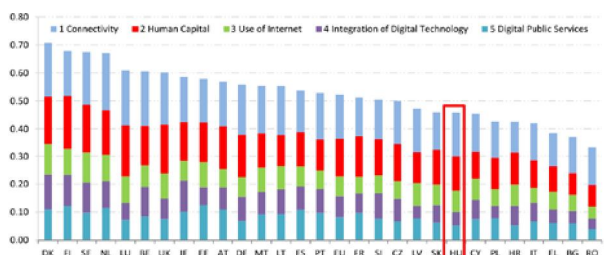

**Figure 1. DESI 2017 country ranking
Source: EC (2017)**

As far as the DESI components concerned, use of Internet is outperform the EU average, and connectivity is in sync with it, however, human capital score is much worse, not to mention integration of digital technology and digital public services, where there is a significant gap between EU and Hungary (Figure 2).

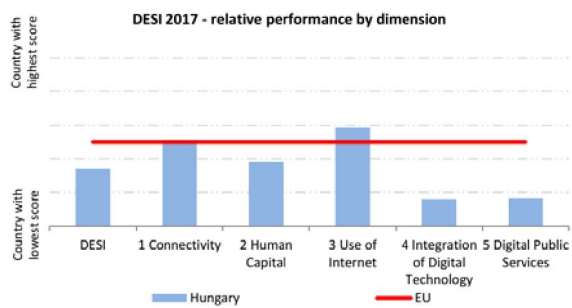

**Figure 2. DESI components in Hungary
Source: EC (2017)**

It is worth investigating the reasons behind the differences in relative performance by dimensions of DESI.

As far as *connectivity* concerned, there is an increase (0.64 in 2017 from 0.60 in 2016), which is chiefly caused by the progress in the take-up and coverage of fast broadband technologies (Table 2.)

**Table 2. Connectivity scores Source: EC (2017)**

| 1 Connectivity | Hungary rank | Hungary score | Cluster score | EU score |
|---|---|---|---|---|
| DESI 2017 | 14 | 0.64 | 0.53 | 0.63 |
| DESI 2016 | 16 | 0.60 | 0.46 | 0.59 |

Hungary ranks 14th in this component, compared to 16th in 2016. Hungary has made progress both in the supply and the demand side. Fast broadband coverage increased to 81% from 78%. The Hungarian government launched two initiatives to increase demand. Primary, a preferential VAT rate is applied to broadband subscriptions in 2017. Secondly, the government created a "digital welfare basic tariff" for non-users. Consequently, a basic broadband package (fixed or mobile) with a 10-15% price discount is available from 2017. 95% of homes in Hungary can now have access to fixed broadband services. 4G coverage is also very high (92%) but mobile broadband penetration is still quite low, only 43 subscriptions per 100 people, which was even lower in the previous period (34 subscriptions per 100 people). This is because mobile broadband prices are significantly higher than in the rest of Europe (Figure 3).

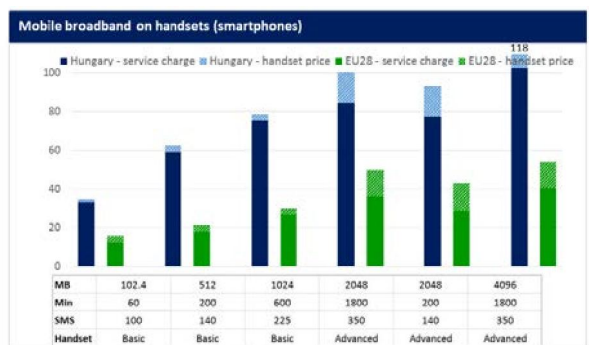

**Figure 3 mobile broadband prices on handsets, Hungary Source: EC (2016)**

As for *Human Capital* component of DESI, Hungary ranks 18th among EU countries slightly below the EU average. Its position in the ranking remained unchanged but progressed 0.05 point in a year time (Table 3.). The country shows a diverse picture in digital skills as only slightly more than the half of the population have at least basic digital skills (51%), whereas ICT specialists represent slightly higher share of the labor force than in EU (3,6% compared to 3.5% in the EU).

**Table 3. Human Capital scores Source: EC (2017)**

| 2 Human Capital | Hungary rank | Hungary score | Cluster score | EU score |
|---|---|---|---|---|
| DESI 2017 | 18 | 0.49 | 0.40 | 0.55 |
| DESI 2016 | 18 | 0.44 | 0.38 | 0.53 |

It's a significant improvement that the number of internet users has been gone up to 78% from 72%.

Telecommunication plays a fundamental role in the family lives of the Hungarian Internet users who are convinced that the more telecommunication device a family uses, the better informed it is. Hungarian users are doing a lot of things on the Internet outstripping the EU average on the use of the internet. Internet makes it easier to get along in life (eNet, 2015). The most popular device in Hungary is the mobile phone including basic mobile phones and smartphones, followed by TV. 96% of the population use mobile phones currently, and 93% have TV. The use of multiple screens (two or three) is on the increase in Hungary. The ratio of STEM (Science, Technology and Mathematics) graduates per 1000 individuals aged 20-29 is quite low (11/1000). In response, the Hungarian Government implemented a new Digital Competences Strategy to increase the ratio of STEM graduates and to address lifelong learning.

The *use of Internet* is the best performing component of DESI in which Hungary scores above the European average (Table 4). On the one hand, Hungarian Internet users are very active in reading news (88% of the individuals who used the Internet in the last 3 months) but they also like using social networks (83%) and making video calls (54%). The use of social networks is the highest here in Europe, outperforming the EU average by 20%. On the other hand, Internet users in Hungary are less engaged in online banking and shopping and only 8% pay for Video on Demand (VOD) services, which is significantly lower than the EU average (21%). It is also unfavorable that the number of online banking users dropped by 2% in a year.

**Table 4. Use of Internet scores Source: EC (2017)**

| 3 Use of Internet | Hungary rank | Hungary score | Cluster score | EU score |
|---|---|---|---|---|
| DESI 2017 | 12 | 0.52 | 0.39 | 0.48 |
| DESI 2016 | 11 | 0.51 | 0.37 | 0.45 |

*Integration of digital technology by businesses* is the biggest problem in Hungary as Hungarian firms should better exploit the possibilities offered by online business, social media and cloud-based applications. With 0.52 points, on the Integration of Digital Technology by businesses, Hungary's ranks 24th, significantly below the EU average (Table 5.). However, Hungary progressed in all indicators and advanced three ranks compared with 2016. Only 13% of enterprises use social media (11% in 2015), 8% send eInvoices (6% in 2015), 8% use cloud services (6% in 2015), 12% of SMEs sell online (10% in 2015) and even less 4.5% sell online cross-border. nevertheless, the e-Commerce turnover went up to 7.6% from 7.0%. The Hungarian government launched two initiatives to boost integration of digital technology by business.

**Table 5. Integration of digital technology Source: EC (2017)**

| 4 Integration of Digital Technology | Hungary rank | Hungary score | Cluster score | EU score |
|---|---|---|---|---|
| DESI 2017 | 24 | 0.24 | 0.27 | 0.37 |
| DESI 2016 | 27 | 0.21 | 0.25 | 0.35 |

The objective of Modern Businesses Programme is to raise awareness, whereas the Programme called "Support of business digital developments" will offer grants and loan financing opportunities.

*Digital public services* are one of the most challenging areas of the digital economy and society in Hungary where there is room for improvement. With 0.35 points Hungary ranks only 27th on this dimension (Table 6.)

**Table 6. Digital Public Services scores Source: EC (2017)**

| 5 Digital Public Services | Hungary rank | Hungary score | Cluster score | EU score |
|---|---|---|---|---|
| DESI 2017 | 27 | 0.35 | 0.43 | 0.55 |
| DESI 2016 | 24 | 0.33 | 0.42 | 0.51 |

Hungary ranks only 23rd on the pre-filled forms, which is the re-use of information across administrations to make life easier for citizens and even worse, 25th on the online service completion. However, the ratio of eGovernment users in Hungary (30%) is not too low considering the low service level of online public service. It is favorable that Nemeslaki, Aranyossy & Sasvári (2016) found high-level of on-line voting intent amongst young Hungarian internet users and that perception of on-line voting would enhance voting desire. Open data is also an issue, where Hungary performed significantly worse - dropped by 7% - compared with the previous year.

All things considered, DESI of Hungary scores above the EU average in the use of internet and is somewhat below the average on connectivity and human capital. Still there are two serious areas to improve: the integration of digital technology by businesses and digital public services. Hungary surpasses the European average in the obtainability and take-up of fast broadband as well as in the use of social networks.

**4. CROSS COUNTRY ANALYSIS OF HUNGARY AND UKRAINE**

After investigating the status of digital economy and society in Hungary, I decided to examine thoroughly the components of DESI by using Consumer Barometer Graph Builder. To make this investigation more interesting, I performed cross country analysis to find out the significant differences between Ukraine and Hungary. Unfortunately, as Ukraine is not a EU member state, Ukraine is not included in DESI. However, Consumer Barometer contains data of Ukraine, too.

81% of Hungarians use the internet as of 2017, which is 1% drop compared with the previous year. However, there was a leap in the percentage of people who access the internet from 73% in 2015 to 82% in 2016. Although the number of Internet users in Ukraine is growing rapidly and steadily, it is still significantly lower (66%) than it was in Hungary five years ago (Figure 4.)

As far as the device use concerned, the mostly used device is the mobile phone in Hungary (96%), while the television in Ukraine. It can be concluded that almost everyone has got a mobile phone and/or a TV set in both countries. The computer, smartphone and tablet penetration are significantly higher in Hungary and the same applies to other digital devices such as digital devices to save or record TV programs, digital devices to stream Internet-content on TV screen, MP3 players, wearable digital devices and eReaders (Figure 5.).

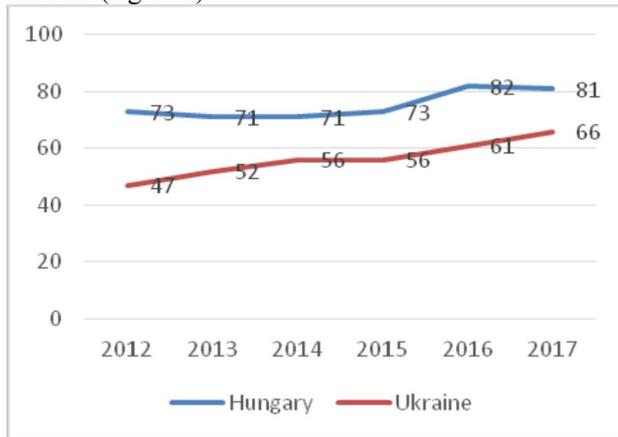

Figure 4. Percentage of people who access the Internet. Source: CB (2017b)

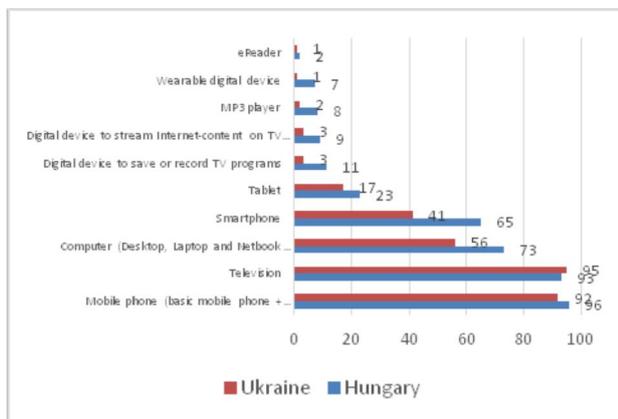

Figure 5. Percentage of people who use a specific device Source: CB (2017c)

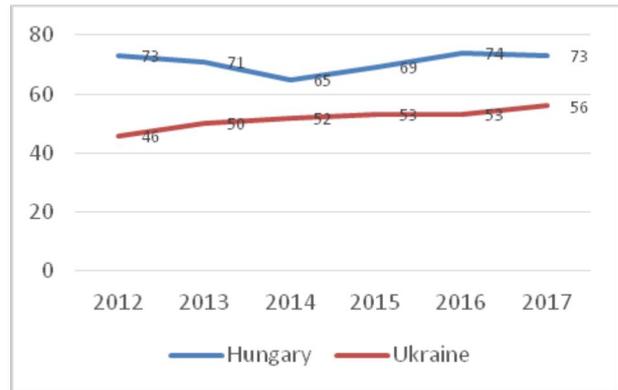

Figure 6. Percentage of people who use a computer Source: CB (2017a)

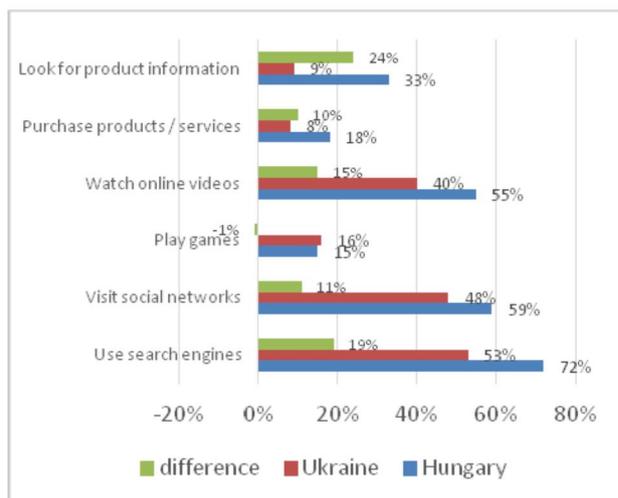

Figure 7. Percentage of people who use a computer for a specific task. Source: CB (2017d)

Currently, 73% of Hungarians use a computer. This is almost the same percentage as of last year. It must be highlighted that there has been a 9% increase in the number of computer users in the last three years as several tasks on the Internet can still be done most effectively by computers. In Ukraine, the number of computer users is slowly increasing, but the usage rate (56%) is still significantly lower than in Hungary (Figure 6).

In Hungary, computers are mainly used for using search engines (72%), visiting social networks (59%) and watching online videos (55%), followed by looking up product information (33%). Playing games on computers (15%) and purchasing product/services (18%) was found to be the least typical activities (Figure 7). Hungary outperformed Ukraine in all but one activities. Playing games in computers is slightly more typical in Ukraine (16%), however there are huge differences in other activities. In Hungary 24% more people look for product information using computers than in Ukraine. The second noticeable difference between the two countries is the search engine use, which is 19% lower in Ukraine. Social networking and online shopping are significantly higher in Hungary. In Ukraine, only 8% of people use a computer for purchasing products and services (Figure 7). Therefore, it can be concluded that eCommerce is still in the birth phase in Ukraine.

The dynamic proliferation of smartphones seems to be unstoppable in Hungary and in Ukraine, too (Figure 8). 65% of Hungarian use a smartphone now, but there was a jump in the percentage from 50% in 2015 to 61% in 2016. 96% of the Hungarians use mobile phone (basic or smart). In Ukraine, the ratio of the smartphone users is 41% and it is growing steadily year by year. Interestingly, it was only 7% five years ago, so the growth is very strong.

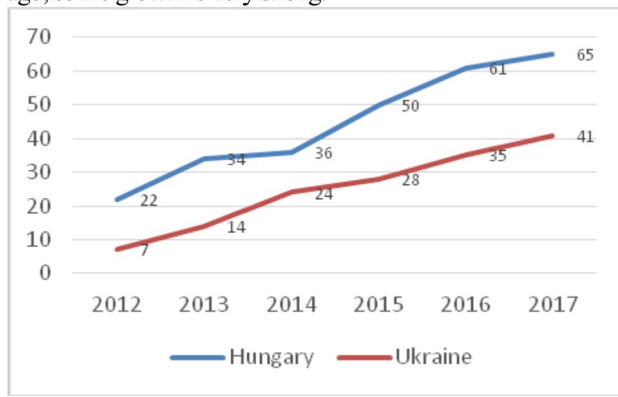

Figure 8. Percentage of people who use a smartphone Source: CB (2017a)

Smartphones have become integral part of the lives of the users in Hungary. Two third of the users never turn off these gadgets, and the majority keep their phones with them all night. Smartphone users are also very active in app consumption, they are mostly interested in free apps, as two thirds of them never download apps for money. The most popular apps are maps and navigation (69%). Apps used for communications (62%), gaming (61%) and social media (58%) have also proved to be indispensable. Most people (77%) usually have been using a smartphone for more than 12 months; while 12% have been using it for less than 6 months. 90% of the users access to the internet through Wi-Fi on their smartphones (CB, 2017).

As far as smartphone functions concerned, in both countries smartphones are mostly used for setting alarm clock, taking photos/videos and telling the time (Figure 9.) Listening to music, checking weather and checking news are also popular daily life activities of smartphone users. It is interesting that significantly higher number of people use their smartphones for reading books and magazines in Ukraine than in Hungary. In general, higher percentage of the population use smartphones for checking news, managing diary and appointments, listening to music and tracking health, diet and activity level in Ukraine. However, Hungarians prefer checking weather, maps and travel information on smartphones.

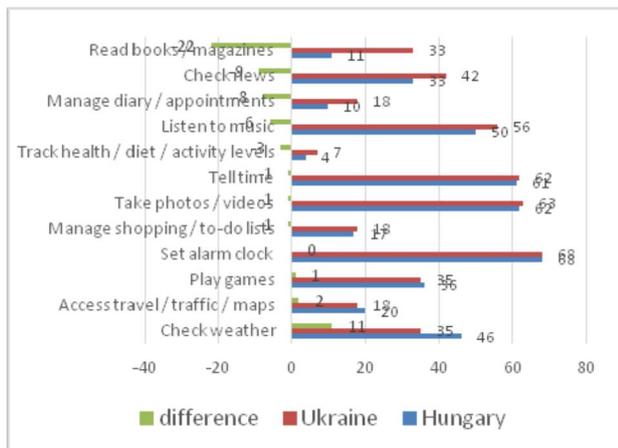

**Figure 9. For what daily life activities do people use their smartphone? Source: CB (2017e)**

The use of tablets is also increasing in both countries. Five years ago, the tablet penetration was insignificant in Hungary and in Ukraine, now the percentage of people using a tablet is 23% in Hungary and 17% in Ukraine. It is also interesting that there has been a boom in the number of tablet users in both countries lately. In Hungary, the number of users doubled in just two years from 2014 to 2016, while in Ukraine there was a 6% increase last year.

The most popular online activities users doing on their tablets at least weekly are: use searching engines (18%), checking emails (15%), visiting social networks (14%) and watching online videos (14%), followed by looking up product information (8%), looking up maps and directions (6%), listening to music (6%). Playing games (3%) and purchasing products/services (1%) are the least typical activities on a tablet (CB, 2016)

As far as multi-screening concerned, percentage of people who do not use any connected device is only 17% in Hungary, which is significantly lower than 34% in Ukraine.

The percentage of people using four, five or more digital devices is also significantly higher in Hungary, so this country has to potential to be a multi-screen country in the near future if this trend continues. On the contrary, Ukraine is in the very early stage of the evolution into a multiscreen nation (Figure 11).

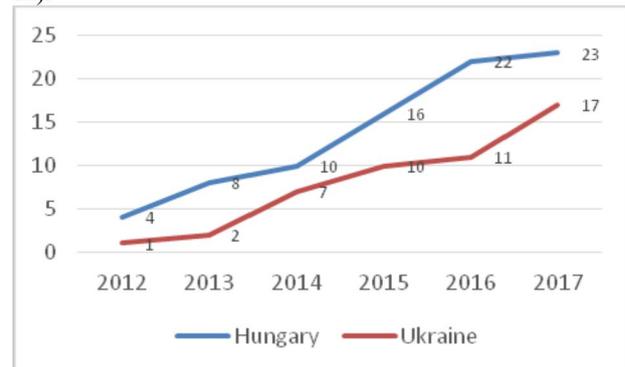

**Figure 10. Percentage of people who use a tablet Source: CB (2017a)**

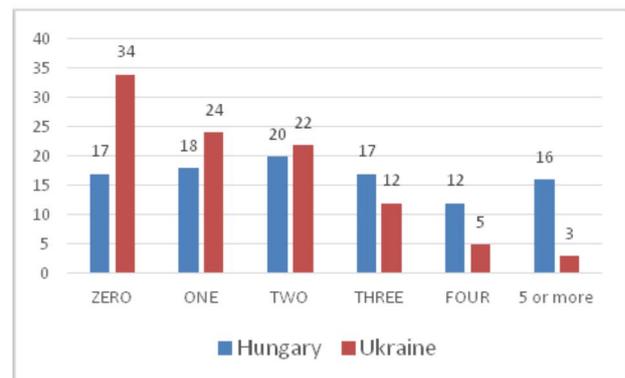

**Figure 11. Percentage of people using the given number of connected devices. Source: CB (2017f)**

### 5. CONCLUSIONS AND OUTLOOK

Based on the above findings it can be concluded that Hungary is more developed in terms of the significant parameters of the digital economy and society than Ukraine. However, compared to other European countries, even Hungary is underdeveloped, emerging country, where the digital society is a bit stronger than the digital economy. Based on my findings, I am convinced that the golden era of Internet related devices will continue as more and more Hungarians and Ukrainians will use such devices and love them more than ever. My findings also support the idea that the DESI index of Hungary is expected to grow further, and Hungary will be soon catching up with the EU average. Considering the high growth rate of Internet, tablet and smartphone penetration in Ukraine, I came to the conclusion that Ukraine will also make significant progress towards a digital society and economy.

In Hungary, the analysis of DESI components justified that both Internet and device usage are on the rise, especially demand for smartphones and broadband mobile services are getting stronger. The use of social networks in Hungary is the highest in Europe and expected to remain unchanged. However, the integration of digital technology by businesses is still a major concern. Internet banking, online shopping and the use of digital public services are the weakest areas that are offering the most opportunities for further development.

Given the obvious similarities and discrepancies, the results of the analysis of the digital development path of Hungary can be used as an input for the digital strategy development in Ukraine.

*Відомості про авторів / Сведения об авторах / About the Authors*

**Szabolcs Nagy**, Associate Professor, University of Miskolc, Faculty of Economics, Institute of Marketing and Tourism; 3515 Miskolc-Egyetemváros, Hungary, +3646/565-111/17-30, nagy.szabolcs@uni-miskolc.hu